\documentclass[%
preprint,
nobibnotes,
amsmath,amssymb,
aps,
pra,
floatfix
]{revtex4-2}

\usepackage{graphicx}
\usepackage{dcolumn}
\usepackage{bm}
\usepackage{hyperref}

\usepackage[caption=false]{subfig}

\begin{document}


\title{The passage of a vortex electron over an inclined grating}

\author{A. Pupasov-Maksimov}
\email{tretiykon@yandex.ru}
 \affiliation{Universidade Federal de Juiz de Fora, Brazil }

\author{D. Karlovets}%
 \email{d.karlovets@gmail.com}
\affiliation{School of Physics and Engineering, ITMO University, 197101 St. Petersburg, Russia
}%

\date{\today}

\begin{abstract}
 We study Smith-Purcell radiation from a conducting grating generated by an inclined passage of a shaped electron wave packet with an electric quadrupole moment in the non-paraxial regime. 
 Spreading of an asymmetric wave packet induces quadrupole corrections to the 
 radiation field. Although the non-paraxial corrections stay small, they are dynamically enhanced during the interaction of the electron with the grating whose length exceeds the Rayleigh length of the packet. 
 To simplify the possible experimental setup where such effects could be measured, we 
 study the dependence of these effects on the inclination angle, i.e. the angle between the mean velocity of the packet and the surface of the grating.
 There is a minimal angle such that the multipole expansion always stays valid at the grating surface. In such a regime, the quadrupole contribution to the Smith-Purcell radiation can become the leading one, which represents a novel quantum effect impossible for classical point-like electrons.
 Thus, the impact of the wave-packet shape (vortex structure or non-spherical shape) can be observed experimentally by comparing the radiation for different 
 orientations of the grating in the single-electron regime.   
\end{abstract}


\maketitle

\section{Introduction}
Many theoretical studies suggest that one can influence the properties of radiation or interactions of a free quantum particle with atoms and condensed matter by adjusting its wave function shape \cite{pan2019spontaneous,zhao2021,hu2017cherenkov,ivanov2013detecting}. 
As experimental capabilities increase, they enable studying vanishingly small interactions between quantum particles, condensed matter and radiation, thus providing reliable experimental tools to test the fundamental theories 
\cite{lankhuijzen1998decay,stobinska2010quantum,peatross2008photoemission,wong2021control}. 

In the present work, we extend our previous studies \cite{karlovets2020non,pupasov2021smith} of the Smith-Purcell radiation 
induced by shaped electron 
wave packets. 
A freely moving wave packet spreads. If such a packet carries a non-vanishing quadrupole momentum, then, according to the Heisenberg equations of motion, it has a quadratic dependence on the evolution time. The standard paraxial regime requires that the evolution time be small compared to the Rayleigh diffraction time.  

We consider the non-paraxial regime of emission since the corresponding corrections become important already at moderate values of orbital angular momentum (OAM) of vortex electrons \cite{karlovets2019dynamical,karlovets2020non}. A vortex electron carries  an orbital angular momentum $\ell\hbar$ with respect to the propagation axis. Such a shaped wave packet is also characterized by multipole moments \cite{karlovets2019intrinsic}.  
These multipole moments can be used to calculate the corrections $dW_{eQ}$ and $dW_{QQ}$ to the intensity $dW_{ee}$ of Smith-Purcell radiation produced by a point charge when the quantum recoil $\eta_q$ is small, 
\begin{eqnarray}
& \displaystyle
\eta_q:= \frac{\omega}{\varepsilon} \ll \frac{dW_{eQ}}{dW_{ee}},\quad 
\frac{\omega}{\varepsilon} \ll \frac{dW_{QQ}}{dW_{ee}},
\label{def:quasi-classical-restrictions}
\end{eqnarray}
and the energy losses stay negligible compared to the electron's energy.
We emphasize that there are two types of quantum corrections to the classical radiation of charge \cite{berestetskii1982quantum,bagrov1993theory,akhiezer1990theory,akhiezer1993semiclassical,baier1968processes}: 
\begin{itemize}
    \item 
The corrections due to recoil, which appear in the operator method \cite{baier1968processes} and the eikonal method \cite{akhiezer1993semiclassical}, 
\item
The corrections due to the finite coherence length of the emitting particle, which appear in the {\it non-paraxial} regime of emission \cite{karlovets2020non}. 
 \end{itemize}
  The latter effects are usually neglected in the conventional paraxial regime, where the packets are almost plane waves. The non-paraxial regime, on the contrary, implies that one cannot neglect the spreading of the packet.
 
 Our previous analytical results \cite{karlovets2020non,pupasov2021smith} highlight three observable effects related to the quadrupole moment and spreading of the wave packet. 
 In contrast to a classical spreading beam, the spreading of the quantum wave packet does not lead to spectral line broadening. Numerical studies of the spectral lines reveal not only an absence of the broadening, but even \textit{a slight narrowing} of the lines due to the charge-quadrupole interference. The quadrupole contribution is dynamically enchanced along the grating, leading to \textit{a nonlinear growth} of the radiation intensity with the grating length. At the same time, the maximum of the radiation intensity with respect to the polar angle is shifted towards smaller angles. 
The magnetic dipole moment results in a small asymmetry of azimuthal distribution both for diffraction and Smith-Purcell radiation \cite{ivanov2013detecting,karlovets2020non,pupasov2021smith}. Nevertheless, the corresponding correction stays small for Smith-Purcell radiation, and we will not consider this contribution here. 

For a direct experimental test, the radiation obtained in various experimental arrangements can be compared: with gratings of different lengths, or with different ratios of the initial mean radius $\bar{\rho}_0$ to the OAM of the packet, $\ell$. However, the experimental task of \textit{isolating small effects on radiation intensity from systematical effects} can be challenging. Instabilities in the wave packet impact parameter, mean velocity $\beta$, various gratings' parameters, etc. could  blur the non-paraxial effects. The usage of a single beam source and a single grating can simplify the possible experimental studies significantly.

\begin{figure}[ht!]
 \centering
 \includegraphics[width=.95\linewidth]{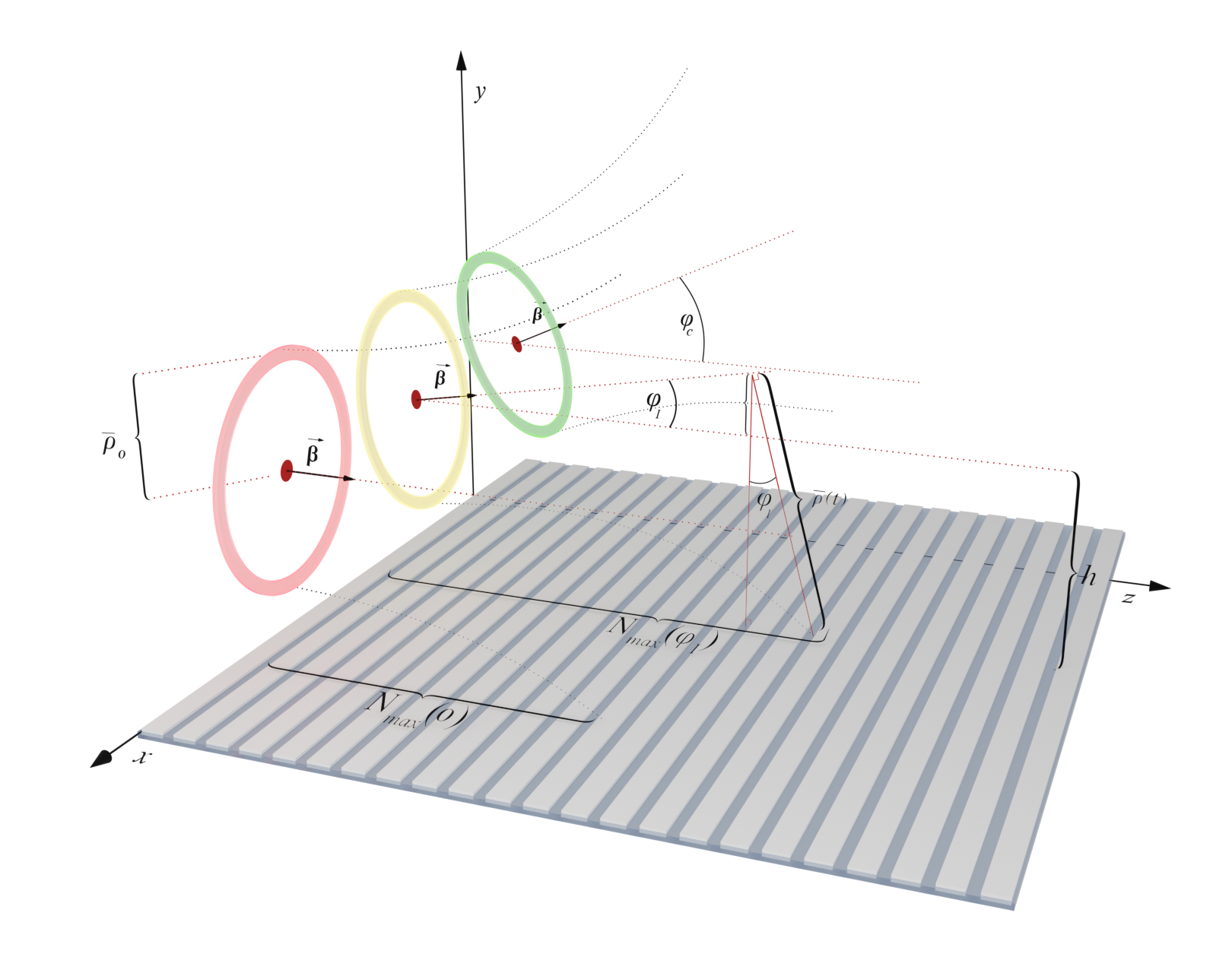}
 \caption{Variation of the inclination angle, i.e. the angle between the mean velocity of the packet and the grating, results in the modification of the interaction length and effective impact parameter. Dashed lines indicate spreading of the wave packets. While the charge contribution (exponentially) decreases with the inclination angle, the quadrupole contribution is increased when $0<\varphi_I<\varphi_c$.}  \label{fig:geometry}
\end{figure}

Therefore, in the present work we consider the case of an inclined passage of a wave packet to provide a theoretical prediction of the dependence of non-paraxial effects on the inclination angle. When the inclination angle $\varphi_I$ is positive (see Fig. \ref{fig:geometry}), motion away from the grating can extend the permitted passage time, and as a result, an increase of non-paraxial contributions can be expected. Let $t_d$ be the diffraction time of the packet. 
Generally, when the inclination angle 
$$\tan \varphi_I>\bar{\rho}_0/(\beta t_d)$$ 
the packet will always stay above the grating. Thus, one can formally consider an inclined passage over an infinite grating. Since the distance between the charge and the grating is increased in this case, the relative intensities of charge and quadrupole contributions can be changed by adjusting the inclination angle. This adjustment can be done simply by rotating the grating without changing the rest of the experimental setup, thus simplifying possible experimental studies.  

Another potential application of our calculations is estimation of the beam spreading effects in the single-electron regime, i.e. a distribution of individual inclination angles and impact parameters of each wave packet. In this case, it is necessary to incoherently average the radiation intensities of the charge and quadrupole contributions over a reasonable distribution of impact parameters and inclination angles, which we expect to be a Gaussian distribution. 

The paper is organized as follows.
First, in section \ref{sec:beams-and-moments}, 
we generalize our considerations for arbitrary axially symmetric wave packets with a quadrupole moment. The corresponding intrinsic quadrupole moment is characterized 
by the function $Q(t)=Q_0+Q_2t^2$, and the constants $Q_0$, $Q_2$ are related to the quadrupole contribution and the non-paraxial corrections. 
Within the multipole expansion, electromagnetic fields of an arbitrary axially symmetric wave packet can be calculated using expressions from \cite{karlovets2019intrinsic}.
This allows us to apply the method of generalized surface currents \cite{karlovets2009generalized,brownell1998spontaneous,potylitsyn2000resonant}, with incident fields obtained by planar rotation (see Fig. \ref{fig:geometry}) of the electromagnetic fields \cite{karlovets2019dynamical}. Calculations of the radiation fields in the wave zone in section \ref{subsec:fourier-integrals} and the spectral and angular distribution of the radiation 
in section \ref{sec:spectral-angular-distr-SPR} follow the line of \cite{pupasov2021smith}. Inclination angle results in the modification of the effective impact parameter and 
appearance of an effective grating length. 
To study angular distributions, we apply numerical integration in the vicinity of the first spectral maximum.

The detailed physical model allows us 
to estimate values of the following parameters: size of the wave packet and its aspect ratio, the orbital angular momentum, velocity, etc., that are compatible with our calculation scheme based on multipole expansion. Note that single-electron regime with a freely propagating packet is realized for low electron currents, much lower than the so-called start current, which is usually less than 100 nA for the electron energy used in a TEM \cite{andrews2005dispersion}.

With obtained analytical expressions, in section \ref{sec:Inclination},
we analyze how the quadrupole contribution can be isolated by changing the inclination angle.
In the Conclusion, we discuss our final experimental proposal for the detection of the quadrupole (non-paraxial) contribution to the Smith-Purcell radiation. Our method is compatible both with moderate OAM values of $\ell\sim 10$ (see also \cite{mcmorran2011electron,zhong2019atomistic,zhong2019high}) and large aspect ratio of Gaussian wave packets.  

Throughout the paper, we use the units with $\hbar=c=|e|=1$.

\section{Wave packets with intrinsic multipole moments \label{sec:beams-and-moments}}
\subsection{Gaussian packets and non-paraxial regime \label{subsec:LG-nonparaxial}}
It is shown in \cite{karlovets2020non} that both the shape and the phase of the electron wave function can affect the transition amplitudes depending on how the final electron is detected. In classical electrodynamics, multipole moments describe electromagnetic fields generated by 
discrete or continuous charge distribution. In particular, the quadrupole moment describes the deviation of the charge distribution from spherical symmetry .

In the works \cite{karlovets2020non,pupasov2021smith}, the generalized LG packet was used to describe a vortex electron with an electric quadrupole moment \cite{karlovets2018relativistic, karlovets2019dynamical} 
\begin{eqnarray}
Q_{\alpha\beta}(t) = \left(\bar{\rho}(t)\right)^2\, \text{diag}\{1/2,1/2,-1\}.
\label{vortex-q-moment}
\end{eqnarray} 
This quadrupole moment increases together with the mean radius of the wave packet 
$$
\bar{\rho}(t)  =\bar{\rho}_0 \sqrt{1 + t^2/t_d^2}.
$$ 

In this work, we generalize our previous results to arbitrary 
wave packets carrying an electric quadrupole moment. 
The simplest wave packet possessing a non-vanishing quadrupole moment is a Gaussian wave packet with different uncertainties
\begin{equation}
\sigma_x\,,\quad \sigma_y\,,\quad \sigma_z\,.
\end{equation}
We isolate the intrinsic quadrupole moment, following \cite{karlovets2019intrinsic}.
Due to the spreading of this free wave packet, its quadrupole moment has a quadratic growth with time
\begin{eqnarray}
Q_{ii}(t) = 2\sigma_i^2-\sigma_j^2-\sigma_k^2+t^2\frac{\lambda_c^2}{4}\left(\frac{2}{\sigma_i^2}-\frac{1}{\sigma_j^2}-\frac{1}{\sigma_k^2}\right)\,,\qquad (i,j,k)=\text{cycle}(x,y,z)\cr
Q_{xy}(t)=Q_{yz}(t)=Q_{xz}(t)=0\,.
\label{General-quadrupole-moments}
\end{eqnarray} 

Note that in the spherically symmetric case, the quadrupole moment vanishes. 
The case of an axially symmetric Gaussian wave packet,  $\sigma_x=\sigma_y=\sigma_\perp$,
is analogous to the non-paraxial Laguerre-Gaussian wave packet, 
and the quadrupole moment tensor reads
\begin{eqnarray}
Q_{\alpha\beta}(t) = Q(t)\, \text{diag}\{1,1,-2\}\,, 
\label{axially-sym-q-moment}
\end{eqnarray} 
\begin{eqnarray}
Q(t)=(\sigma_\perp^2-\sigma_z^2)\left(1-\frac{\lambda_c^2t^2}{4\sigma_\perp^2\sigma_z^2}\right)=Q_0+Q_2t^2
\label{q-polynomial}
\end{eqnarray}
Moreover, using the symmetry considerations and Heisenberg equations for the quadrupole moment, one can find that \eqref{axially-sym-q-moment}, \eqref{q-polynomial} are valid for 
arbitrary axially symmetric free wave packets.
As a result, in this case, all the calculations of Smith-Purcell radiation can be made just by substituting $\bar{\rho}_0^2\to Q_0$ and $\ell^2\frac{\lambda_c^2}{\bar{\rho}_0^2}\to Q_2$.  

Note that an axially symmetric wave packet without OAM implies that $Q_0Q_2<0$, while 
in the case of LG packet, we have $Q_0>0$, $Q_2>0$. 
For the LG packet, there is a special case when the packet has 
the same diffraction time in all directions.
This vortex packet spreads, and its transverse area is doubled during the diffraction time $t_d$,
\begin{eqnarray}
t_d = \frac{m \bar{\rho}_0^2}{|\ell|} = \frac{t_c}{|\ell|}\, \left (\frac{\bar{\rho}_0}{\lambda_c}\right)^2 \gg t_c,
\label{def:diffraction-time}
\end{eqnarray}
which is large compared to the Compton time scale $t_c = \lambda_c/c \approx 1.3\times 10^{-21}\,\text{sec.}$, $\lambda_c \approx 3.9\times 10^{-11}\, \text{cm}$. Moreover, the proportions between its transverse and longitudinal dimensions stay unchanged.   
The quadrupole moment \eqref{vortex-q-moment} follows the same law (it is doubled during the diffraction time), and as a result, it increases monotonically.

For a Gaussian wave packet, such a uniform spreading requires  spherical symmetry of the packet, thus leading to a vanishing quadrupole moment. In general, the packet will invert its aspect ratio during the spreading. As a result, $\text{sign}\, Q_0\neq \text{sign}\, Q_2$, and the components of the quadrupole moment change their sign.

\subsection{Geometrical restrictions}
When the LG packet moves nearby the grating, its finite spreading time \eqref{def:diffraction-time} puts an upper limit on the possible impact parameter $h$, on the initial mean radius of the packet and on the grating length for a given inclination angle $\varphi_I$. The geometry (see Fig. \ref{fig:geometry}) yields the following inequality
\begin{align}\label{uneq:collision-condition}
\bar{\rho}(t)\cos\varphi_I <\beta t \sin\varphi_I+h    \,,
\end{align}
which can be solved to find the maximum passage time
\begin{equation}
    t<t_{max}(\phi_I)=t_d\frac{\beta t_d h \sin\varphi_I+\bar{\rho}_0 \cos\varphi_I \sqrt{h^2+ \bar{\rho}_0^2\cos^2\varphi_I+\beta^2t_d^2\sin^2\varphi_I}}{\bar{\rho}_0^2\cos^2\varphi_I-\beta^2t_d^2\sin^2\varphi_I}.
    \label{def:max-passage-time}
\end{equation}

Note that the singularity of the denominator defines the critical angle 
\begin{align}
\tan\varphi_c=\frac{\bar{\rho}_0}{\beta t_d}\,,    
\end{align}
and the passage time is unlimited for $\varphi_I>\varphi_c$. Therefore, in the following text, the estimates on the maximum time and maximum grating length are considered only when $\varphi_I<\varphi_c$.

 
The maximum number of strips $N_{max}(\varphi_I)$ is
\begin{eqnarray}
\displaystyle
N_{max}(\varphi_I)=\beta t_{max}/d\,, \qquad N_{max}(\varphi_I>\varphi_c)=\infty\,.
\label{N-max-expression}
\end{eqnarray}

Note that in the experiment, the number of strips can be larger then $N_{max}$, and when $N>N_{max}$, the electrons may collide with the grating. However, the grating transition radiation, which is produced by electron collisions with the grating,
 should not affect the radiation pattern at larger polar angles. Therefore, we take 
$N_{max}(\varphi_I)d$ as the formation length for the Smith-Purcell radiation when the inclination angles obey $\varphi_I<\varphi_c$.

The geometry implies that $\bar{\rho}_0<h=\bar{\rho}(t_{max})$, or $\bar{\rho}_0\ll h$ for a long grating. In practice, only low diffraction orders can be considered, so that $d\sim\beta\lambda$ for the emission angles $\Theta \sim 90^{\circ}$. 
In the case of a parallel passage, a rough estimate of the maximum number of strips for $h \approx h_{eff} \sim 0.1 \lambda, \beta \approx 0.5, \bar{\rho}_0 \sim 1$ nm yields \cite{pupasov2021smith}
\begin{eqnarray}
& \displaystyle
N_{max}(\varphi_I=0)\lesssim 
\frac{h\bar{\rho}_0}{|\ell| \lambda \lambda_c }\sim \frac{10^3}{|\ell|}.
\label{N-limitations}
\end{eqnarray}
Hence, if $N_{max}(\varphi_I=0) \gg 1$, then $|\ell| < 10^3$. 


\subsection{Electric fields of the quadrupole moment}
\begin{figure}[ht!]
 \centering
 \includegraphics[width=.95\linewidth]{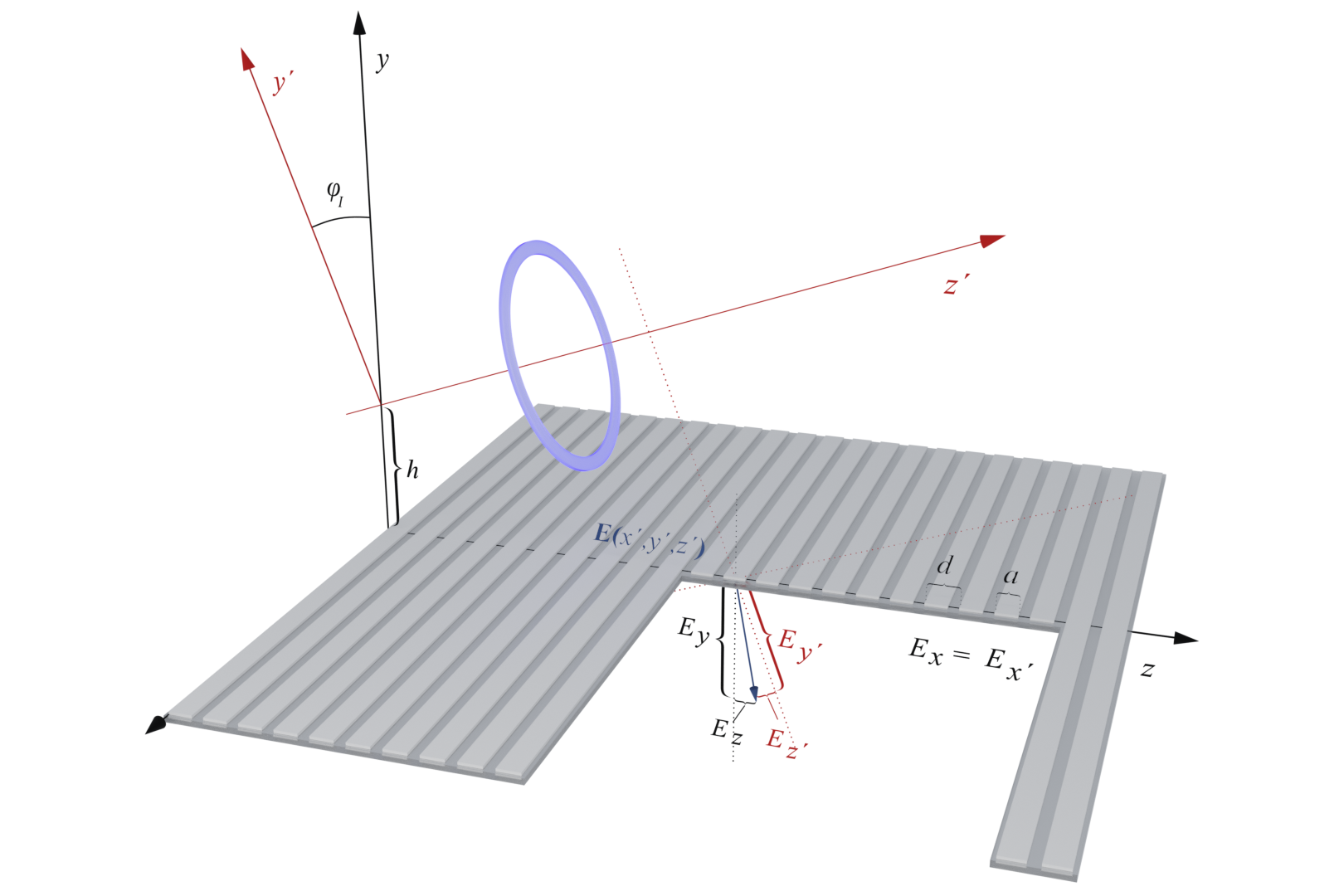}
 \caption{Inclined passage of a Laguerre-Gaussian (LG) packet with respect to the conducting grating with the period $d$ and width of the strip $a$. The electric fields of the packet induce surface currents on the grating, which are the source of the radiation. The transformation from $(x',y',z')$ to $(x,y,z)$ coordinate system, and the transformation of the electric field components are given in \eqref{def:coordinate-transformation}, \eqref{def:field-transformation}} 
 \label{fig:coordinate-systems}
\end{figure}

The physical mechanism of Smith-Purcell radiation is related to the polarization and 
surface currents induced by the electromagnetic interaction of a passing charge with the grating \cite{karlovets2009generalized,potylitsyn2000resonant}. To calculate the surface current, we need only the electric fields.  
In the Cartesian coordinates $x', y', z'$ aligned with the vector of the mean velocity of the wave packet, the electric field of the quadrupole moment 
can be calculated 
as follows  \cite{karlovets2019dynamical} :
\begin{flalign}
& {\bm E}_{Q}({\bm r'}, t) = \frac{\gamma}{4R'^3} {\bm \rho'}\Bigg (3\frac{Q_0}{R'^2} \left(1 - 5\frac{R_{z'}^2}{R'^2}\right) + Q_2\Big[3 \frac{T_{z'}^2}{R'^2}\left(1 - 5\frac{R_{z'}^2}{R'^2}\right) + 3 \frac{R_{z'}^2}{R'^2} - 6\beta \frac{R_{z'} T_{z'}}{R'^2} -  1\Big]\Bigg) + \cr 
& \frac{\gamma}{4R'^3}({\bm z'} - {\bm \beta}t) \Bigg (3\frac{Q_0}{R'^2} \left(3 - 5\frac{R_{z'}^2}{R'^2}\right) + Q_2 \Big[ \frac{T_{z'}^2}{R'^2}\left(3 - 5\frac{R_{z'}^2}{R'^2}\right)+ 3 \frac{R_{z'}^2}{R'^2} - 1\Big]\Bigg).
\label{EQlabsep}
\end{flalign}
where the following notations are used: ${\bm \beta} = \left(0, 0, \beta\right),\, {\bm z'} = \left(0, 0, z'\right)$,
\begin{align}
 {\bm \rho}' = \{x',y'\},\quad R_{z'}:=\gamma(z' - \beta t),\quad T_{z'}:=\gamma(t-\beta z'),\cr
 {\bm R}' = \{{\bm \rho}', \gamma (z' - \beta t)\} \,,\quad \gamma=(1-\beta^2)^{-1/2}
\label{notations-R-Rz-Tz}
\end{align} 
Note the complex time dependence of the quadrupole field in a fixed point of grating. First, the time dependence is included in the change of the distance between the packet and a fixed point. Second, there is an additional time dependence due to 
the terms proportional to $T_{z'}$, $T_{z'}^2$.

The transformation from the accompanying coordinate frame to the laboratory frame (see also Fig. \ref{fig:coordinate-systems}) reads 
\begin{flalign}
& x' =  x\,,\cr
& y' =  (y-h)\cos(\varphi_I)-z\sin(\varphi_I)\,,\cr
& z' =  z\cos(\varphi_I)+(y-h)\sin(\varphi_I)\,.
\label{def:coordinate-transformation}
\end{flalign}
The components of the electric field in the two coordinate systems are related by the rotation \begin{flalign}
& E_x = E_{x'}\,,\cr
& E_y = E_{y'}\cos(\varphi_I)+E_{z'}\sin(\varphi_I)\,,\cr
& E_z = E_{z'}\cos(\varphi_I)-E_{y'}\sin(\varphi_I)\,.
\label{def:field-transformation}
\end{flalign}

Following the generalized surface current model developed in Ref. \cite{karlovets2009generalized}, the radiation fields at large distances are written as 
\begin{equation}
{\bm E}^R\approx \frac{i\omega {\rm e}^{i kr_0}}{2\pi r_0}\int {\bm e}_0  \times \left[\,{\bm n} \times {\bm E} (k_x,y,z,\omega) \right]{\rm e}^{-ik_z z}dz\,,
\end{equation}
where the electric field ${\bm E}={\bm E}_e+{\bm E}_Q$ is a sum of the Coulomb 
field of the charge and the electric field of the quadrupole moment.
In the case of Smith-Purcell radiation, the integration is performed along the periodic grating.
This expression is obtained for the far-field region, where the curl can be substituted by ${\bm e}_0$. 
The current density induced by the incident field
\begin{equation}
\label{def:generalized-current}
{\bm j}(w)=\frac{1}{2\pi}\, {\bm e}_0 \times \left[\,{\bm n} \times {\bm E}(w)\right]\,,
\end{equation}
is defined by 
a vector product of ${\bm E}$, i.e. the electromagnetic field of the electron incident on the surface of an ideally conducting grating, a normal to the surface ${\bm n}$ and the unit vector
$$
{\bm e}_0=\frac{{\bm r}_0}{|{\bm r}_0|}=\left(\sin \Theta \cos \Phi,\sin \Theta \sin \Phi,\cos \Theta \right).
$$
For moderate electron energies required for the observation of the effects discussed in this work, the normal component of the surface current is crucial: for this reason, we employ the model of Ref.\cite{karlovets2009generalized}, which is more general than that of Brownell et al. \cite{brownell1998spontaneous}.

\subsection{Surface currents and radiation field \label{subsec:radiation-fields}}
When the Fourier transform of $xt$ is performed (see Appendix \ref{subsec:fourier-integrals} and Eq.\eqref{E-quadrupole-z-polynomial}), the surface current density
$$
{\bm j}=\frac{1}{2\pi}\,\left(-E_x e_{0y},\, E_x e_{0x}+E_z e_{0z},\, -E_z e_{0y}\right)\,,
$$
reads
\begin{equation}
\label{def:generalized-current-structure}
{\bm j}(k_x, y', z',\omega) = 
\exp \left( i z'\frac{\omega}{\beta} \right)
\left({\bm j}_0(k_x, y',\omega)+{\bm j}_{Q_1}(k_x, y',\omega)\,z'+{\bm j}_{Q_2}(k_x, y,\omega)\,z'^2\right),
\end{equation}
where $y'$ and $z'$ should be substituted via \eqref{def:coordinate-transformation}. 
The first term contains generalized currents, which are uniform along the grating, ${\bm j}_{0}(k_x, y',\omega)={\bm j}_e(k_x, y',\omega)+{\bm j}_{Q_0}(k_x, y',\omega)$, while the next terms are related to the linear and quadratic time-dependent quadrupole contributions.

The integration with respect to the $z$-coordinate along the periodic grating 
\begin{equation}
\int\limits_0^{N d}dz\left({\bm j}_{0}+{\bm j}_{Q_1}\,z+{\bm j}_{Q_2}\,z^2\right)
\exp \left[ i z \left(\omega \Theta_I+i\sin(\varphi_I)\mu \right)\right]=
\label{integral-for-radiation-field}    
\end{equation}
$$
=\left({\bm j}_{0}+{\bm j}_{Q_1}(-i\partial_{\Omega_I})+{\bm j}_{Q_2}(-i\partial_{\Omega_I})^2\right)F_I(\Omega_I)\,,
$$
can be performed using the derivatives of the following \textit{generating function}
\begin{align}\label{def:gratting-form-factor}
&  F_I(\Omega_I)=\sum_{j=0}^{N}\int\limits_{jd}^{jd+a}dz 
    \exp \left( i z \left(\Omega_I+i\sin(\varphi_I)\mu \right)\right)=\cr
&    \frac{2\sin(\frac{a(\Omega_I+iM_I}{2})}{(\Omega_I+iM_I)}\frac{\sin\left(\frac{N d\left(\Omega_I+iM_I \right)}{2}\right)}{\sin\left(\frac{d \left(\Omega_I+iM_I \right)}{2}\right)}
\exp\left(\frac{(i\Omega_I-M_I)}{2}(a + (N - 1)d)\right),
\end{align}
with respect to the parameter $\Omega_I$, where
\begin{eqnarray}
\Theta_I = \frac{\cos(\varphi_I)}{\beta} - \cos(\Theta)\,,\qquad 
\Omega_I=\omega \Theta_I\,,\quad M_I=\sin(\varphi_I)\mu=\sin(\varphi_I)\sqrt{\frac{\omega^2}{\gamma^2\beta^2}+k_x^2}
\end{eqnarray}
Due to the inclined path of the wave packet, the singularity of the denominator in \eqref{def:gratting-form-factor} shifts to the complex frequency 
\begin{align}
\omega=\left(\frac{2\pi g}{d}+i    \sin(\varphi_I)\sqrt{\frac{1}{\gamma^2\beta^2}+\sin^2\Theta\cos^2\Phi}\right)\left( \frac{\cos(\varphi_I)}{\beta} - \cos(\Theta)\right)^{-1}
\end{align}
Its real part corresponds to the resonance peaks 
\begin{align}
\omega_g = \frac{2\pi}{\lambda_g} =\frac{2\pi\, g}{d}\left( \frac{\cos(\varphi_I)}{\beta} - \cos(\Theta)\right)^{-1} ,\ g=1,2,3,...
\end{align}
which is slightly shifted compared to the case of the parallel passage.
We can estimate the width of the spectral lines as follows. When the passage is parallel, 
the width
\begin{equation}
\Gamma=\frac{\delta \omega}{\omega}=\frac{1}{N}    
\end{equation}
is related to the finite grating length. In other words, the integration of $\exp(iz\omega)$ is performed over a finite interval. Alternatively, one can integrate over the whole semi-infinite line introducing a small imaginary shift  $\exp(iz(\omega+i\Gamma))$. Since the inclination results in such a small imaginary shift, we obtain the spectral width 
\begin{align}
\Gamma_I=\frac{1}{N}+\sin(\varphi_I)\sqrt{\frac{1}{\gamma^2\beta^2}+\cos^2\Phi\sin^2\Theta}
\end{align}
Hereinafter, the effective number of strips is equal to 
\begin{align}\label{def-Neff}
N_{eff}=\frac{1}{\Gamma_I}=\frac{N}{1+N\sin{\varphi_I}\sqrt{\frac{1}{\gamma^2\beta^2}+\cos^2\Phi\sin^2\Theta}}    
\end{align}

When the inclination angle is positive, the grating can be considered as infinite, and the corresponding limit is written as:
\begin{align}\label{def:gratting-form-factor-inf}
& F_{I,\infty}(\Theta_I)= \lim\limits_{N\to\infty} F_I(\Theta_I)=\frac{1}{(M_I-i\Omega_I)}\frac{\exp\left( (i\Omega_I-M_I)a\right)-1}{\exp\left( (i\Omega_I-M_I)d\right)-1},
\end{align}
In the following text, these compact notations are used: 
$$
\partial_{\Omega_I}F_I(\Omega_I)=F'_I(\Omega_I),\quad \partial^j_{\Omega_I}F_I(\Omega_I)=F^{(j)}(\Omega_I).
$$ 

\section{Multipole corrections to the Smith-Purcell radiation from the LG wave packet \label{sec:spectral-angular-distr-SPR}}
\subsection{Spectral distribution of the Smith-Purcell radiation from the LG-wave packet \label{subsec:spectral-distr-SPR}}

The distribution of the radiated energy over the frequencies and angles,
\begin{equation}
\frac{d^2W}{d\omega d\Omega} = r_0^2|{\bm E}^R|\,,
\end{equation}
is a sum of the following terms:
\begin{align}
& \label{expr-dWee-by-FF}
dW_{ee}= \frac{\omega^2}{4\pi^2}|{\bm j}_e|^2|F_I(\Omega_I)|^2\,,\\
& \label{expr-dWeQk-by-FF}
dW_{eQ_k}=i^k\frac{\omega^2}{4\pi^2}\left[{\bm j}_e{\bm j}_{Q_k}^*F_I(\Omega_I)F_I^{(k)}(\Omega_1)^*+(-1)^k{\bm j}_e^*{\bm j}_{Q_k}F_I(\Omega_1)^*F_I^{(k)}(\Omega_1)\right]
\,,\\
& \label{expr-dWQQ-by-FF}
dW_{Q_jQ_k}=\frac{(-i)^{j+k}\omega^2}{4\pi^2}\left[(-1)^k{\bm j}_{Q_j}{\bm j}_{Q_k}^*F_I^{(j)}(\Omega_I)F_I^{(k)}(\Omega_I)^*+(-1)^j{\bm j}_{Q_j}^*{\bm j}_{Q_k}F_I^{(j)}(\Omega_I)^*F^{(k)}(\Omega_I)\right].
\end{align}
Then, the radiation of a charge with inclined passage over an infinite grating is
\begin{align}\label{expr-dWee-incl-inf}
&
\frac{d^2W_{e,I\infty}}{d\omega d\Omega}=
\exp\left(-\frac{\omega h_I}{\beta\gamma}\sqrt{1+\beta^2\gamma^2 \cos^2\Phi \sin^2\Theta}\right)\cr
&
\times
\frac{\left(\cos(a \omega \Theta_I) - \cosh\left(a \omega \sin(\varphi_I) \sqrt{
    1/(\beta^2 \gamma^2) +\cos^2\Phi \sin^2\Theta}\right)\right)}{\left(\cos(d \omega \Theta_I) - \cosh\left(d \omega \sin(\varphi_I) \sqrt{
    1/(\beta^2 \gamma^2) +\cos^2\Phi \sin^2\Theta}\right)\right)}\left(1+\beta^2\gamma^2 \cos^2\Phi \sin^2\Theta\right)^{-1}\cr
&
\times
[\left(\cos^2(\Theta)+\sin^2(\Phi)\sin^2(\Theta)\right)(\cos^2(\varphi_I)+\gamma^2 \sin^2(\varphi_I)) +
\cr
&
\beta \gamma^2 \cos^2(\Phi) \sin^2(\Theta)
\left(2\cos(\varphi_I)\cos(\Theta)+ \beta \gamma^2 \cos^2(\Theta) \sin^2(\varphi_I) + 
 \beta \gamma^2 \sin^2(\Theta)(1 + \sin^2(\Phi) \sin^2(\varphi_I))\right)]\cr
& 
\times \left( \gamma^2 \cos^2(\varphi_I) - 2 \beta \gamma^2 \cos(\varphi_I) \cos(\Theta) + 
 \beta^2 \gamma^2 \cos^2(\Theta) + 
\sin^2(\varphi_I) (1+\beta^2\gamma^2 \cos^2\Phi \sin^2\Theta)\right)^{-1}\,,\cr
\end{align}
Because of the inclination, the impact parameter is included in the exponential factor in combination with the strip length $a$ and period $d$  
\begin{eqnarray}
h_I=2h \cos(\varphi_I) + (a - d) \sin(\varphi_I)\,.
\end{eqnarray}
One can see that the angular dependence of $d^2W_{e,I\infty}$ becomes $d^2W_{e}$  when the inclination angle is zero. When $a=d$, our result coincides with the case of radiation diffraction on a conducting semi-plane \cite{karlovets2011theory}.


The quadrupole contributions from ${\bm j}_{Q_1}$ and ${\bm j}_{Q_2}$ are defined by the real parts of the product of currents, and by the form factor $F_I$ and its derivatives. Nevertheless, explicit calculations show \footnote{See the code in the public repository \cite{pupasov2019git}}  that these terms also have a
factorized structure 
\begin{equation}\label{expr-dWeQ12-factorized-structure}
dW_{eQ_j}=\exp\left(-\frac{\omega h_I}{\beta\gamma}\sqrt{1+\beta^2\gamma^2 \cos^2\Phi \sin^2\Theta}\right)
P_{eQ_j}(k_x,y,\omega)F_{eQ_j}(\omega,k_z).
\end{equation}

\begin{figure}[h!]
\centering
   \includegraphics[width=.95\linewidth]{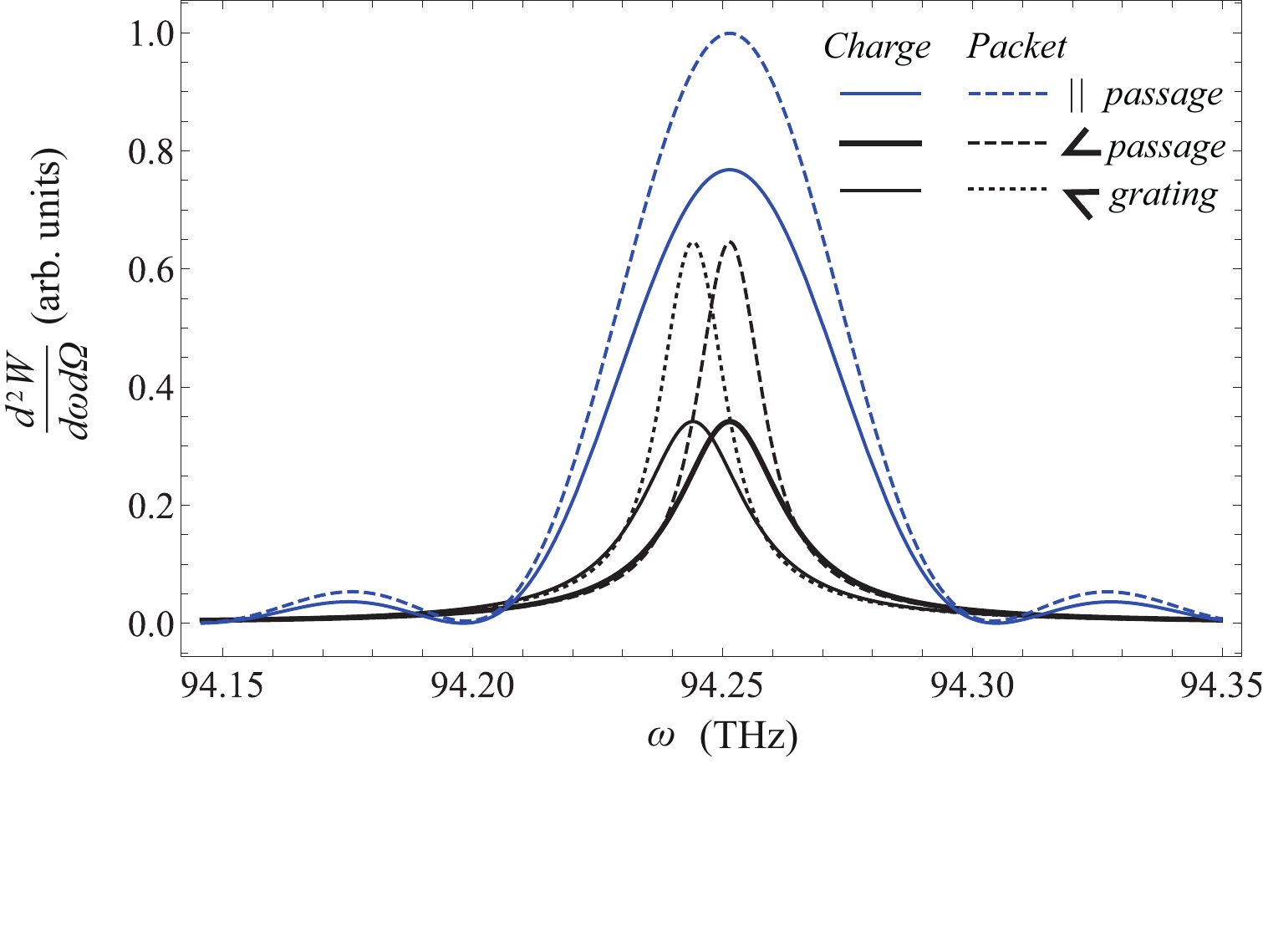} 
    \caption{Comparison of the radiation spectrum of an ordinary electron (solid curves) and of a vortex electron packet (dashed curves) for the parallel passage and the inclined passage with a critical inclination $\varphi_I=0.0045\deg$ defined by the impact parameter $h=2.8\mu m$, $\bar{\rho }_0= 100\text{nm}$, $\ell=10$, $\beta=0.5$. 
 Effects of beam and grating inclinations are compared.
 A zenith direction perpendicular to the grating plane
 $\Theta=\Phi=\frac{\pi}{2}$ is considered. 
 The grating period $d=0.01\,$ mm, $a=d/2$, and the number of strips $N=3500$. }
    \label{fig:spectral-curve-1mm}
\end{figure}

Here, the functions $P_{eQ_j}(k_x,y,\omega)$ define the angular distributions, and $F_{eQ_j}(\omega,k_z)$ determine the positions of the spectral lines and their width (therefore, we will call $F_{eQ_j}(\omega,k_z)$ {\it a spectral factor}).
In Fig. \ref{fig:spectral-curve-1mm}, we compare the intensities of radiation from the charge, the quadrupole, and from their interference. It can be seen that the radiation from the wave packet decreases more slowly compared to the charge radiation (or a wave packet with no quadrupole moment) when the inclination is increased. The spectral line of the wave-packet radiation is also slightly sharper.

For parallel passage, the charge and the charge-quadrupole contributions have almost the same azimuthal dependence \eqref{expr-dWee-SP}, which is defined mostly by the exponential factor for $\beta\sim 0.5$. In the case of an inclined passage, these azimuthal dependencies should be (linearly) mixed after the linear transformation of the fields and the coordinate transformations. However, since they all have the same form (for those parameters that we have already investigated), this mixing will have little effect on the azimuthal distributions.

\subsection{Qualitative analysis and multipole expansion \label{subsec:qualitative-analysis}}

We suppose that the axially symmetric wave packet is shaped to have only a non-vanishing electric quadrupole moment, similar to the case of \eqref{vortex-q-moment} (a vortex wave-packet) or  \eqref{axially-sym-q-moment} (Gaussian wave packet). In this case, the total radiation intensity $dW$ includes classical radiation from the point charge $dW_{ee}$, the interference term $dW_{eQ}$ and the radiation from quadrupole $dW_{QQ}$:
\begin{equation}\label{def:radiation-intensity-interference-complete}
\frac{dW}{d\omega d\Omega}\equiv dW=dW_{ee}+dW_{eQ}+dW_{QQ}
\end{equation}
In the case of a Gaussian wave packet, higher-order multipole moments vanish. For Laguerre-Gaussian packets, higher-order multipole moments are small compared to the dipole and quadrupole ones, and this formula can be considered an approximation. 
This perturbative regime was derived from QED in the quasi-classical approximation when the final electron is detected in the vortex state, \cite{karlovets2020non}.

For parallel passage, we neglected the $dW_{QQ}$ contribution because of the finite 
grating length. In the case of infinite grating, an inclined passage with a valid multipole expansion is possible, therefore, we should keep this term.  
In the case of inclined passage, all considerations of \cite{pupasov2021smith} are still valid. We only need to substitute $N_{max}$ by $N_{eff}$ from \eqref{def-Neff}. For a large $N_{max}$ or for inclination angles larger than the critical one, $\varphi_I>\varphi_c$, when $N_{max}=\infty$,
we can estimate $N_{eff}=\beta\gamma/sin(\varphi_I)$ .  

Using the explicit formulas \eqref{expr-dWeQk-by-FF} (see also \cite{pupasov2021smith}), one can distinguish three different corrections from the charge-quadrupole interference: $dW_{eQ_0}$, $dW_{eQ_1}$, and $dW_{eQ_2}$. Their relative contributions are:
\begin{eqnarray}
&& \displaystyle
\frac{dW_{eQ_0}}{dW_{ee}} \sim \eta_{Q_0}:= \frac{\bar{\rho}_0^2}{h_{\text{eff}}^2} - \text{quasi-classical quadrupole contribution},
\label{def:eQ0-interf-term-parameter}\\
&& \displaystyle
\frac{dW_{eQ_{1}}}{dW_{ee}} \sim \eta_{Q_1}:= \ell^2  \frac{\lambda_c^2}{\bar{\rho}_0^2} - \text{ordinary non-paraxial contribution \cite{karlovets2018relativistic}},
\label{def:eQ1-interf-term-parameter}\\
&& \displaystyle
\frac{dW_{eQ_{2}}}{dW_{ee}} \sim  \eta_{Q_2}:= N_{eff}^2\, \ell^2 \frac{\lambda_c^2}{\bar{\rho}_0^2}  - \text{dynamically enhanced non-paraxial contribution \cite{karlovets2019dynamical}},
\label{def:eQ2-interf-term-parameter}
\end{eqnarray}
where the effective impact parameter of Smith-Purcell radiation naturally appears 
\begin{equation}
\ h_{\text{eff}} = \frac{\beta\gamma\lambda}{2\pi} = \frac{\beta\gamma}{\omega} \sim 0.1\, \lambda\ \text{for}\ \beta \approx 0.4-0.8.
\label{def:effective-impact-parameter}
\end{equation}

It follows from \eqref{def:eQ2-interf-term-parameter} that a large effective number of strips $N_{eff} \gg 1$ can lead to \textit{the non-paraxial regime of emission} with 
\begin{eqnarray}
&& \displaystyle \eta_{Q_1} \ll 1,\ \eta_{Q_2} \lesssim 1,
\label{paramineq}
\end{eqnarray}
when the quadrupole contribution becomes noticeable. 

All the inequalities and restrictions from \cite{pupasov2021smith} can be summarized as follows. \textit{The packet radius should be smaller than the wavelength of the emitted radiation}.  This is just a condition of the multipole expansion in the wave zone.
In this case, we can also safely neglect the possible effects of wave packet reduction due to the measurement of the electron position by observing an emitted photon.

Using the definition  for  the  quantum  and  classical  regimes  by  Renieri  \cite{renieri1984free}, such a wave packet should be considered within the “classical regime”:{\it The    electron    operates    in   the     finite-length     homogeneously     broadened   classical  regime   if  throughout   the  entire   interaction   length   L   its   axial   position   can   be   quantum   mechanically  localized  with  an  accuracy  better  than  the  wavelength  of the  electromagnetic  force  wave  with  which  it  interacts}  \cite{friedman1988spontaneous}. However, despite the fact that the electron is well-localized, the internal structure of the wave packet can be seen in the radiation intensity (\cite{karlovets2020non}, Figs. 3 and 4). Modifications of the radiation intensity are created by the quadrupole moment, which is an entirely quantum characteristic of the wave packet. Therefore, we call this modification a non-linear quantum effect.    

In the case of an efficiently infinite grating, the energy losses of the electron may become noticeable, or its size can become larger than the wavelength emitted during its effective path $N_{eff}d$ over the grating.
From the width of the spectral line we estimate 
$$N_{eff}=\frac{\gamma \beta}{\sin{\varphi_c}}=\frac{\gamma\beta^2t_d}{\bar{\rho}_0}$$
Then, during the effective interaction time, the vortex wave packet will spread 
to 
$$
\bar{\rho}\left(\frac{N_{eff}d}{\beta}\right)=\bar{\rho}_0\sqrt{1+\gamma^2\beta^2\frac{d^2}{\bar{\rho}_0^2}}
$$
The resulting size should be small compared to the emitted wavelength 
$$
\bar{\rho}_0^2<\lambda^2-\gamma^2\beta^2d^2\,.
$$
This condition is definitely violated when the right-hand side of the inequality is negative. This occurs when  
 $\beta\geq 0.58$ (forward radiation) and $\beta\geq 0.78$ (vertical radiation).
However, in this case $N_{eff}$ is large, and very long gratings are required to fulfill this condition.

The baseline parameter set for the IR and THz radiation for parallel passage \cite{pupasov2021smith} is
\begin{itemize}
\item 
$\bar{\rho}_0=10^0-10^2\, \text{nm}$, $\ell \sim 10^1-10^3$, $N \lesssim 10^0-10^3$,   
\end{itemize}

\section{Radiation intensity from a wave packet and a point charge as a function of the inclination angle \label{sec:Inclination}}
\begin{figure}[!ht]
\centering
\subfloat[Subfigure 1 list of figures text][$\ell=10$, $h=0.92\, \mu \text{m}$]{
\includegraphics[width=0.62\textwidth]{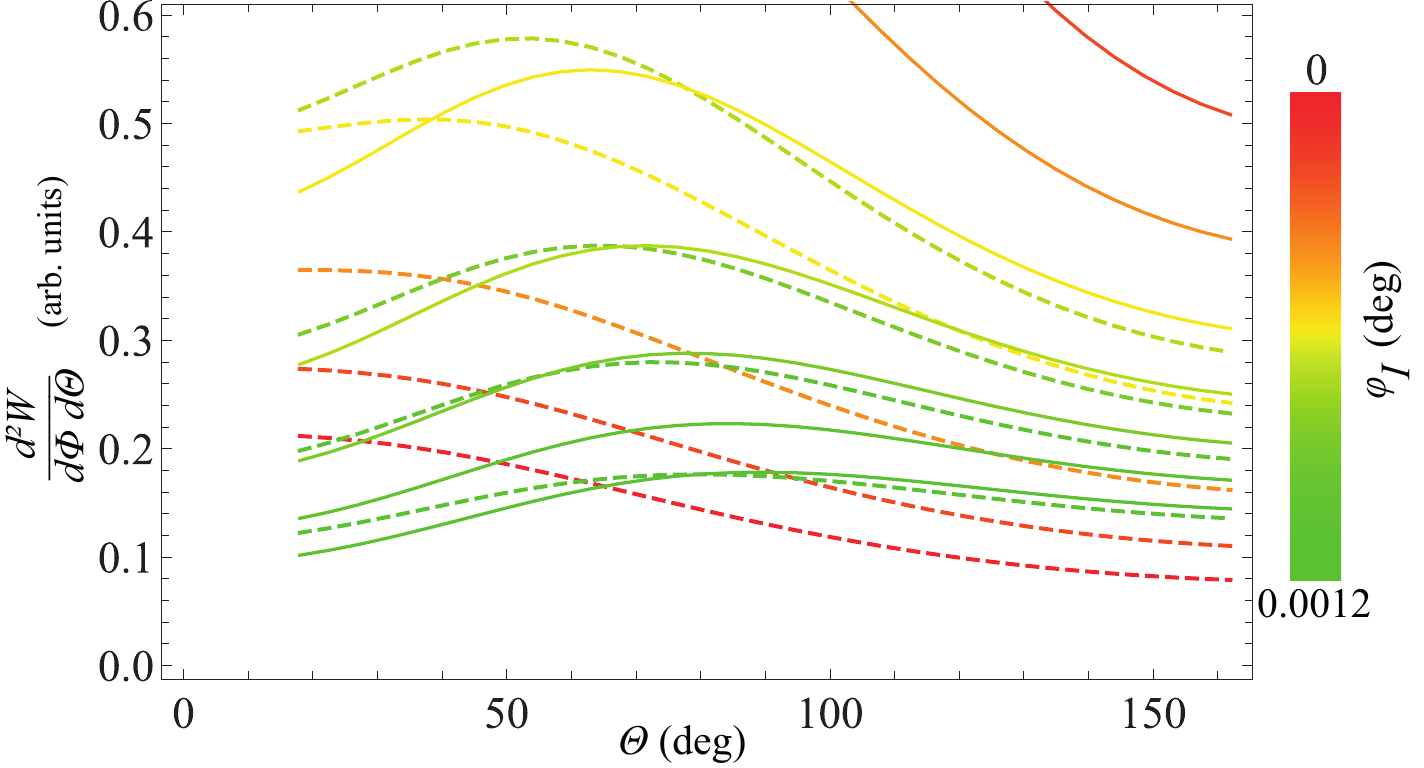}
\label{fig:phi-near}}
\\
\subfloat[Subfigure 3 list of figures text][$\ell=10$, $h=2.76\, \mu \text{m}$]{
\includegraphics[width=0.62\textwidth]{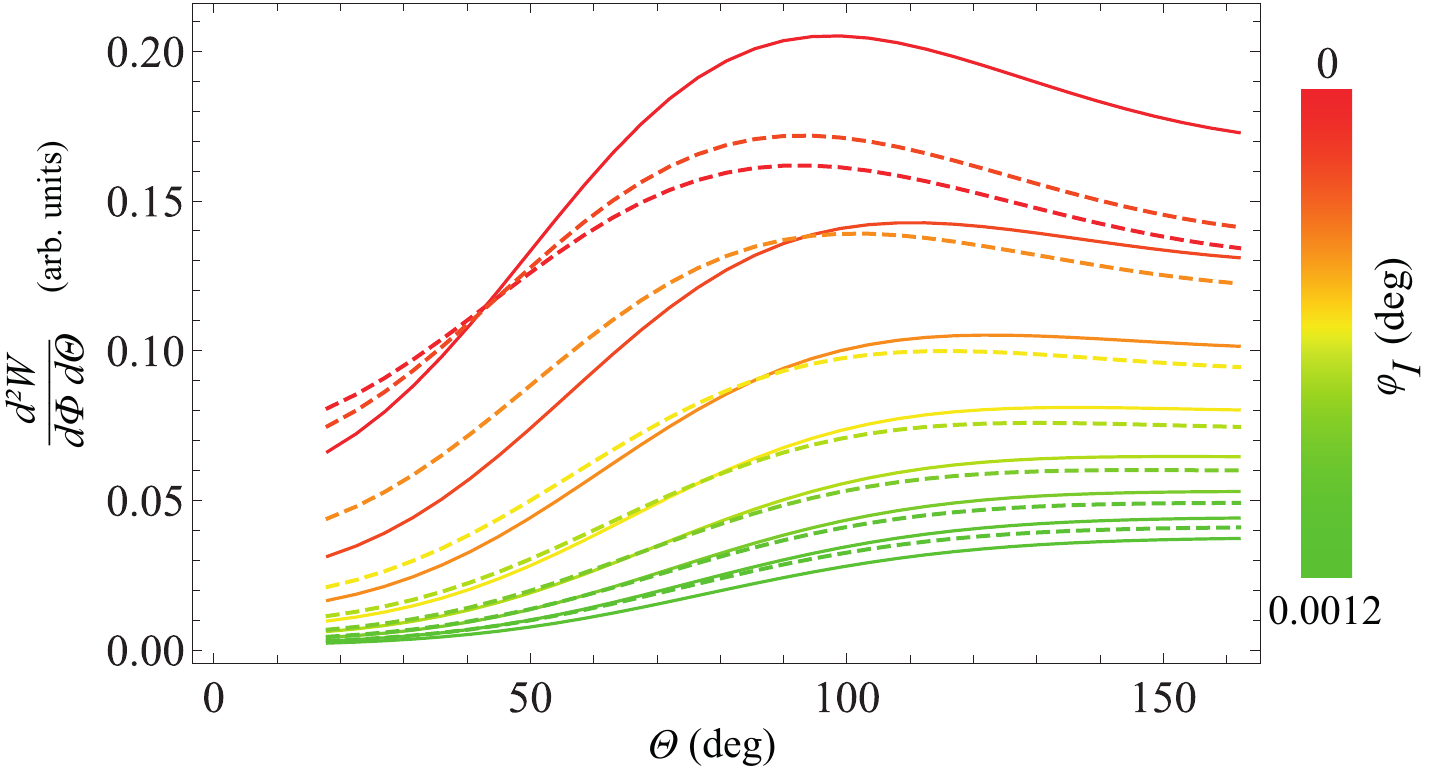}
\label{fig:phi-opt}}
\\
\subfloat[Subfigure 3 list of figures text][$\ell=10$, $h=9.2\, \mu \text{m}$]{
\includegraphics[width=0.62\textwidth]{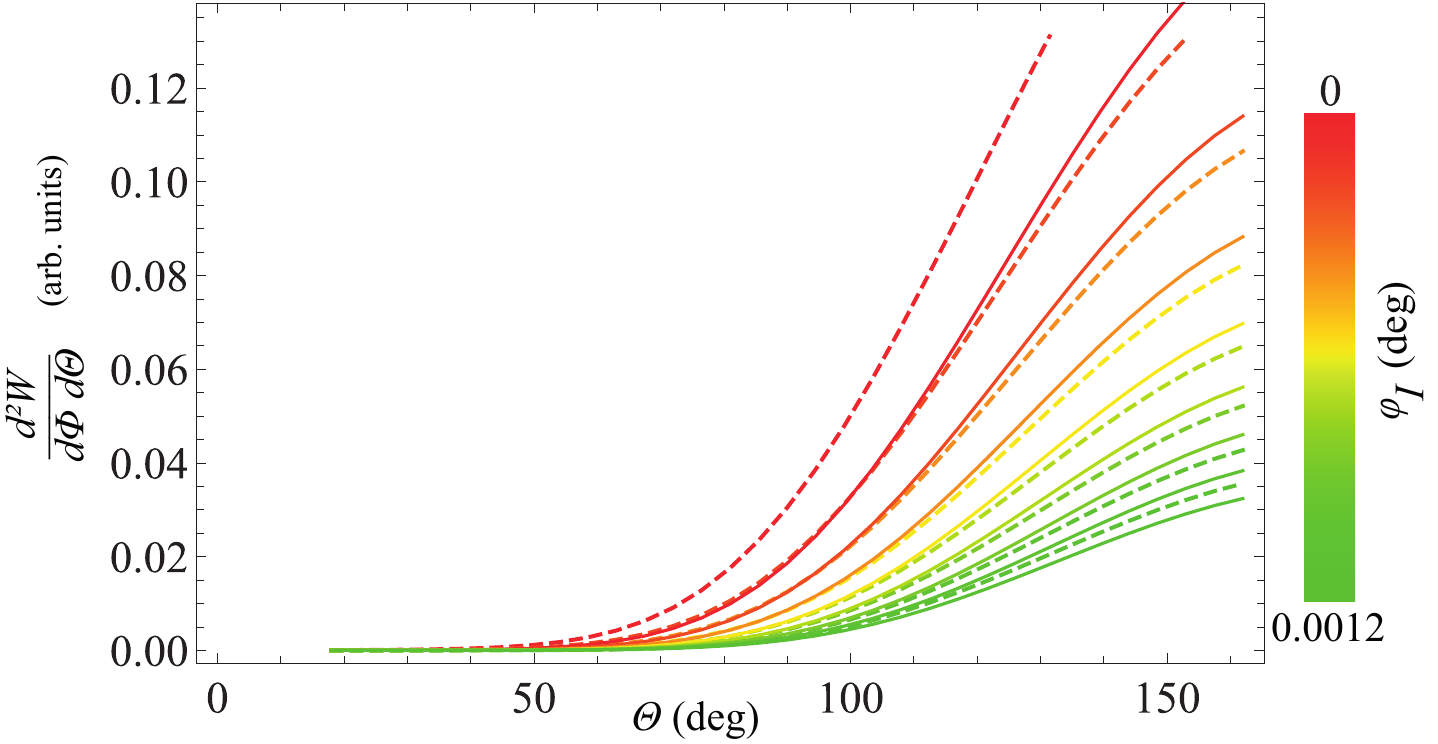}
\label{fig:phi-far}}
\caption{Polar dependence of radiation intensity in the case of a charge (solid lines) and a wave packet (dashed lines) when the inclination angle varies form 0 to $1.5\varphi_c$ (illustrated with change of color from red to green, see the color bar). The number of strips $N_{max}(\varphi_I=0)=1800$ for the parallel passage of the wave packet corresponds to the impact parameter $h=2.8\,\mu\text{m}$, velocity $\beta=0.5$, period $d=0.01\,$ mm, OAM $\ell=10$, initial mean radius $\bar{\rho}_0=50\,$nm, and grating length $2.5$ cm.}
\label{fig:polar-dependence}
\end{figure}

To compare the radiation intensity from the wave packet and the point charge in a fixed direction, we need to integrate $dW$ with respect to the frequency.
At a non-zero inclination angle, analytical integration of quadrupole terms seems to be impossible. However, numerical integration converges rapidly and allows visualization of the resulting angular distributions. We also test our numerical integration in the case of parallel passage, comparing it with the analytical angular distributions of the charge radiation in the main diffraction order $g=1$ 
\begin{eqnarray}\label{expr-dWee-SP}
&& \displaystyle
\frac{dW_{ee}}{ d\Omega}=N
\frac{d^2 \omega_1^3}{\pi^2}\sin^2\left(\frac{a\pi}{d}\right)
\exp\left(-\frac{2\omega_1 y}{\beta\gamma }\sqrt{1+\beta^2\gamma^2 \cos^2\Phi \sin^2\Theta}\right)
\cr
&& \displaystyle 
\times\frac{\cos^2\Theta+2\beta\gamma^2\cos^2\Phi\cos\Theta\sin^2\Theta+\sin^2\Phi\sin^2\Theta+\beta^2\gamma^4\cos^2\Phi\sin^4\Theta}{\beta^2\gamma^2(1+\beta^2\gamma^2 \cos^2\Phi \sin^2\Theta)}\,,
\end{eqnarray}
which is valid for large $N$.

Since the polar distribution strongly depends on the impact parameter, in Fig. \ref{fig:polar-dependence} we compared the evolution of polar distributions for varying inclination angles for different impact parameters. There is a competition between the dynamical enhancement of the quadrupole contribution and the general exponential decay of radiation intensity that occurs with increasing impact parameter or inclination angle. 
It can be seen that the polar distribution for wave packets is shifted to the smaller polar angles compared to the polar distribution of charge radiation, and it behaves differently under variations of inclination angle. This confirms our idea that \textit{one can distinguish radiation patterns from a wave packet and a point charge} by varying the inclination angles. The polar dependence of the radiation intensity may be a suitable candidate for experimental measurements. First, at the critical inclination angle, the change in the polar dependence is not small (see Figure \ref{fig:polar-dependence}). 

\begin{figure}[!ht]
\centering
\subfloat[Subfigure 1 list of figures text][$\ell=10$, $h=0.92\, \mu \text{m}$]{
\includegraphics[width=0.4\textwidth]{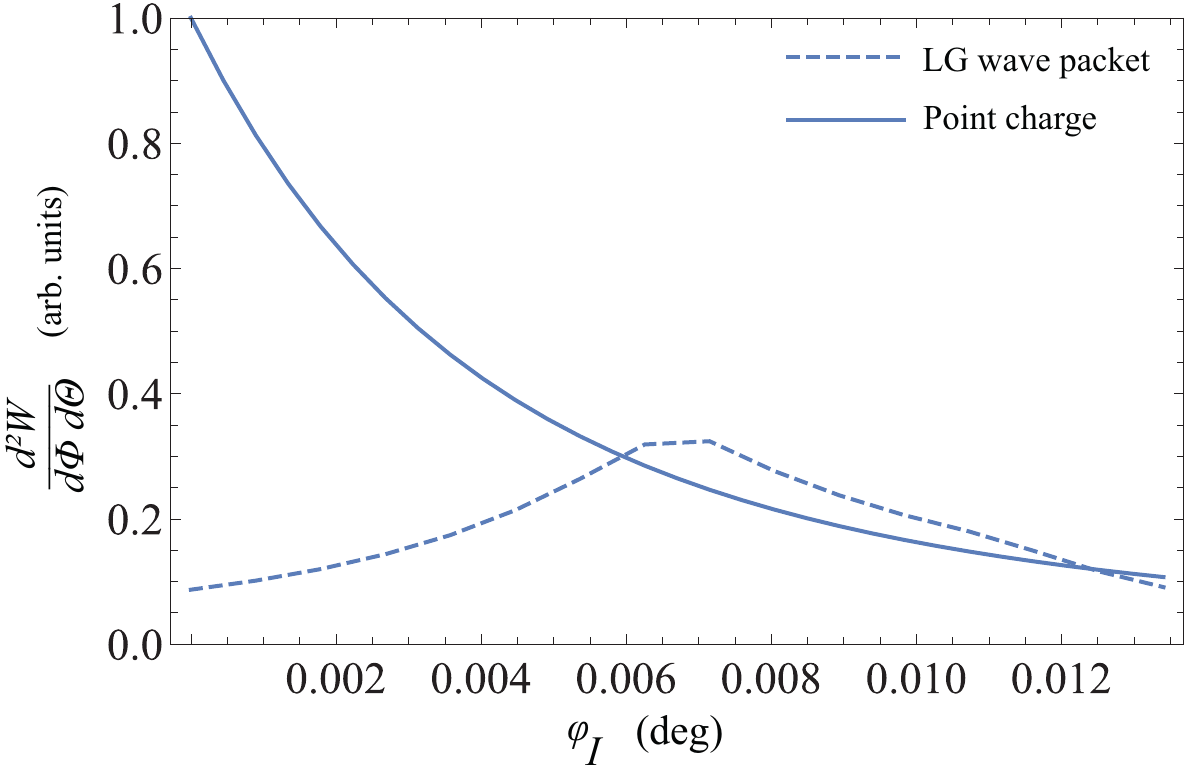}
\label{fig:phi-near-detect}}
\qquad
\subfloat[Subfigure 2 list of figures text][$\ell=100$, $h=0.92\, \mu \text{m}$]{
\includegraphics[width=0.4\textwidth]{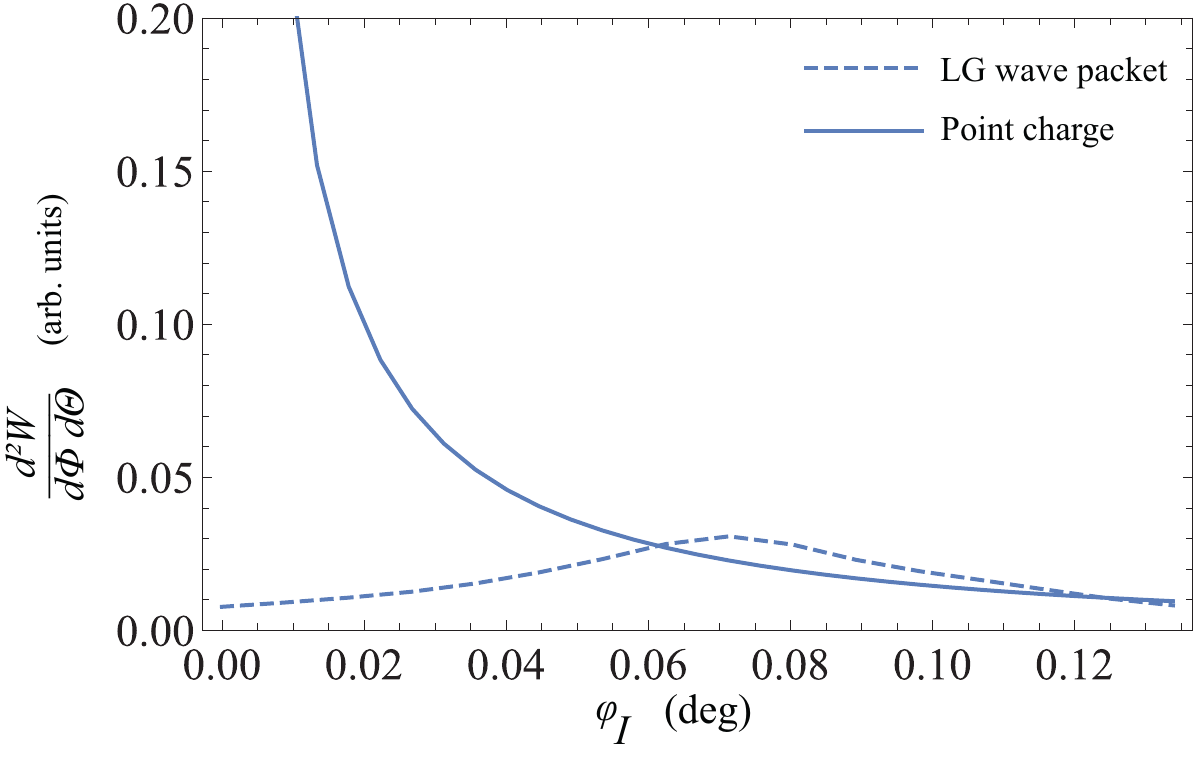}
\label{fig:phi100-near-detect}}
\\
\subfloat[Subfigure 3 list of figures text][$\ell=10$, $h=2.76\, \mu \text{m}$]{
\includegraphics[width=0.4\textwidth]{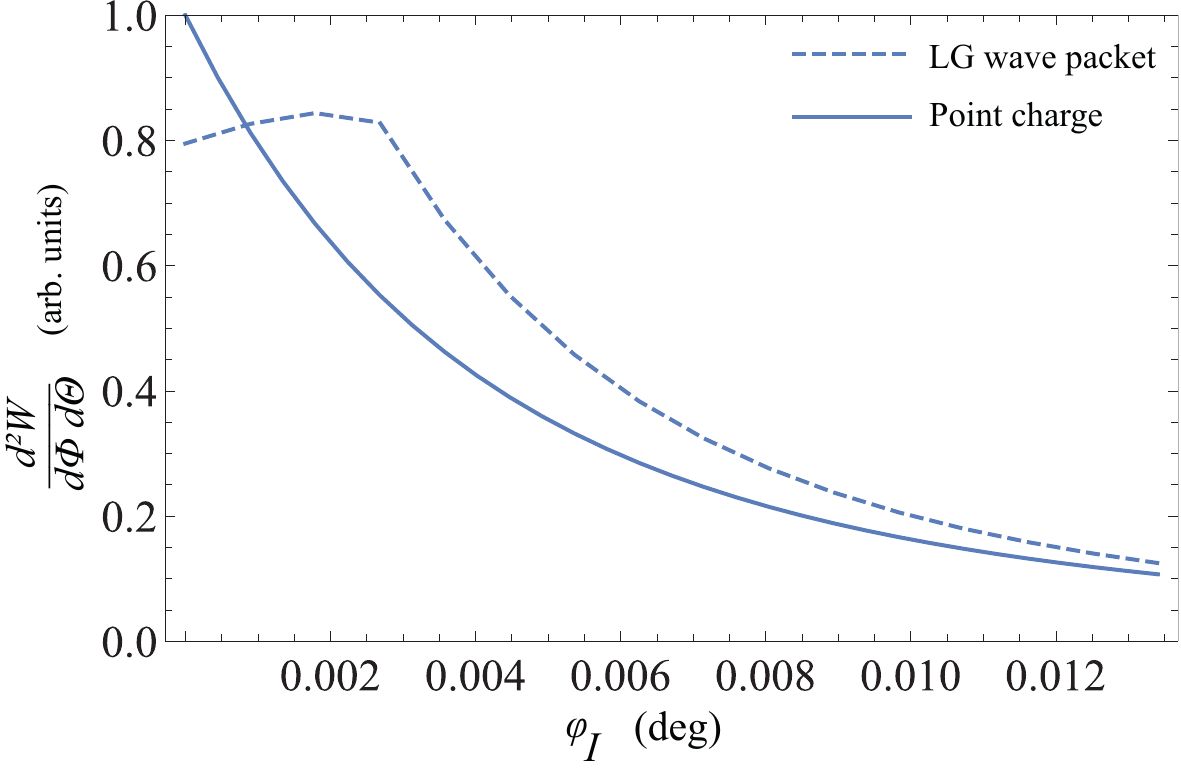}
\label{fig:phi-opt-detect}}
\qquad
\subfloat[Subfigure 4 list of figures text][$\ell=100$, $h=2.76\, \mu \text{m}$]{
\includegraphics[width=0.4\textwidth]{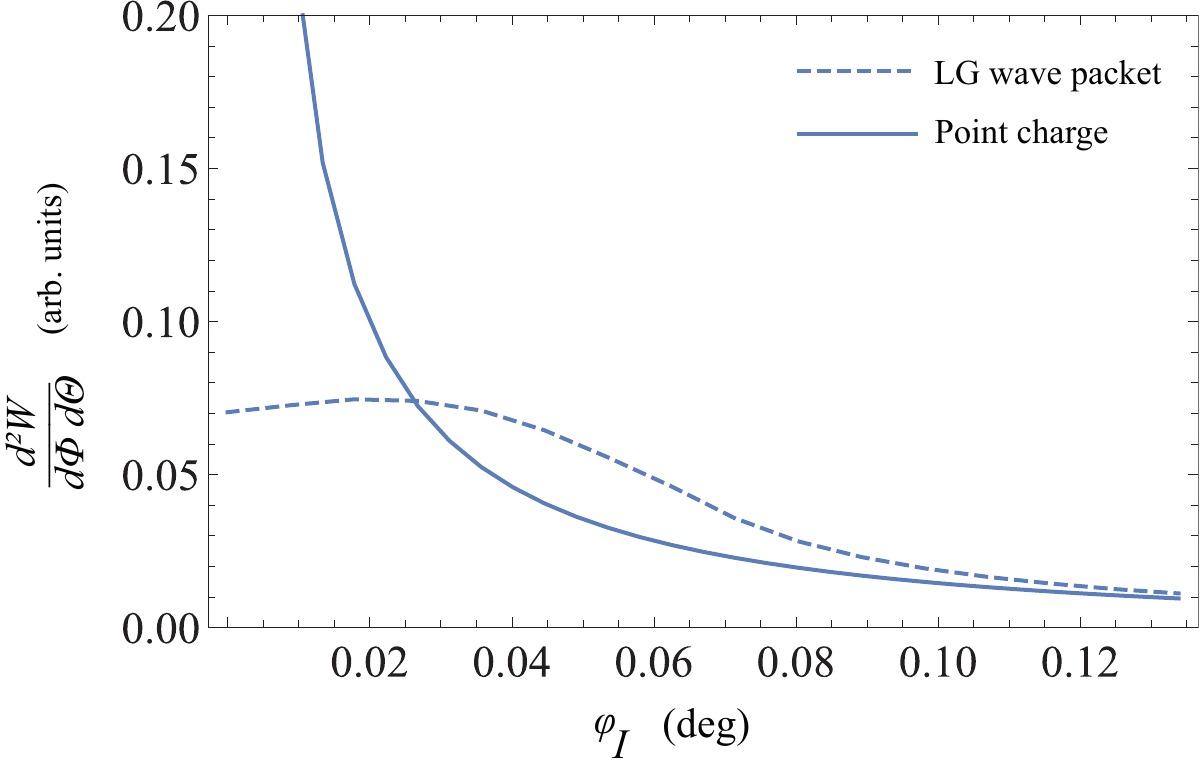}
\label{fig:phi100-opt-detect}}
\\
\subfloat[Subfigure 3 list of figures text][$\ell=10$, $h=9.2\, \mu \text{m}$]{
\includegraphics[width=0.4\textwidth]{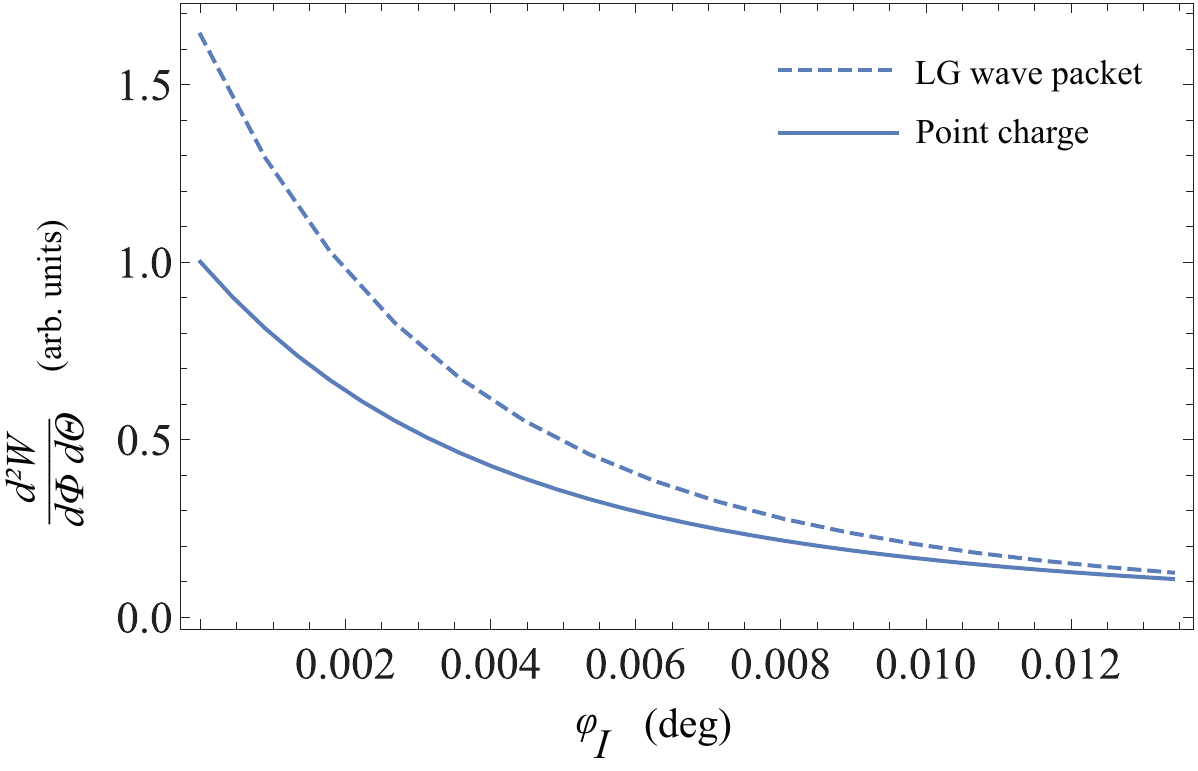}
\label{fig:phi-far-detect}}
\qquad
\subfloat[Subfigure 4 list of figures text][$\ell=100$, $h=9.2\, \mu \text{m}$]{
\includegraphics[width=0.4\textwidth]{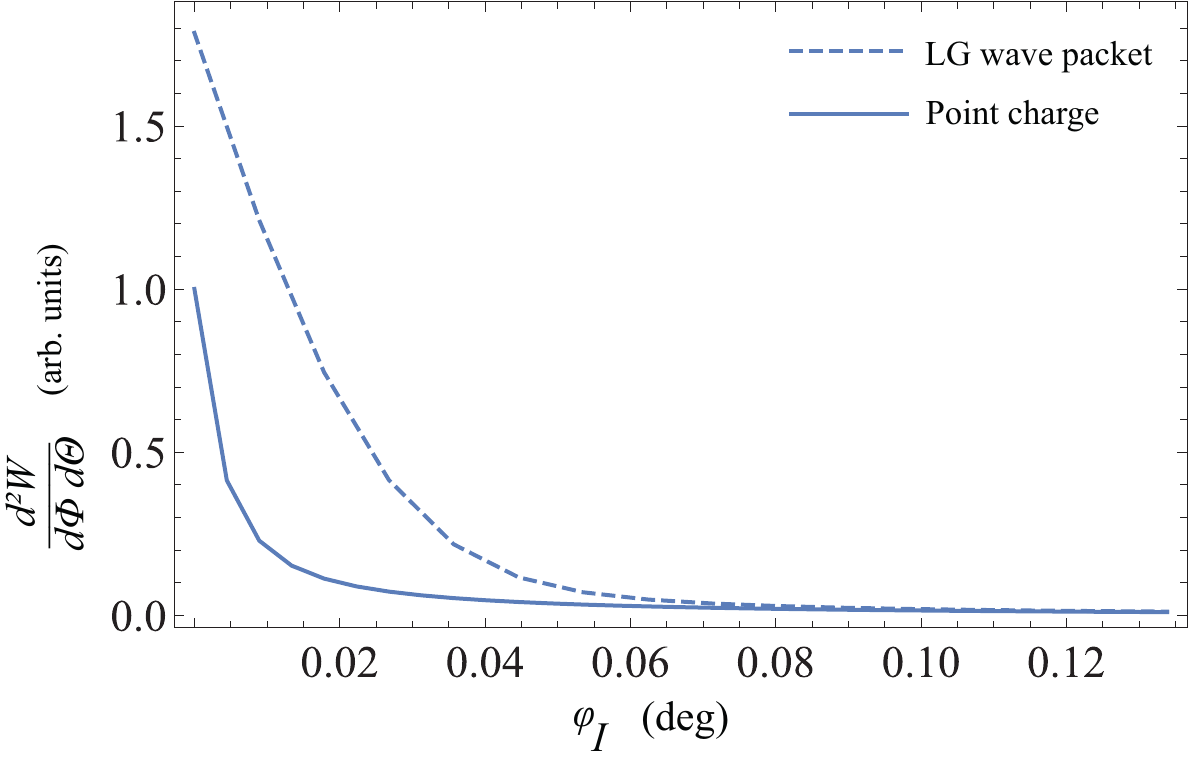}
\label{fig:phi100-far-detect}}
\caption{Radiation intensity in vertical direction $\Theta=\Phi=\frac{\pi}{2}$ as a function of the inclination angle $\varphi_I$ for an LG wave packet with OAM, compared to the radiation intensity from a point charge.  The effective impact parameter is $h_{eff}=1.84\,\mu\text{m}$, number of strips $N=2500$, velocity $\beta=0.5$, period $d=0.01\,$ mm, OAM $\ell=10$, and the initial mean radius $\bar{\rho}_0=50\,$nm.}
\label{fig:inclination-dependence}
\end{figure}

In the Fig. \ref{fig:inclination-dependence} we fix a detector in the zenith direction 
$\Phi=\Theta=\pi/2$ and observe how the radiation intensity changes with variations of inclination angle. Radiation from the charge demonstrates exponential decay with increasing inclination angle. In contrast, for the wave packet, initial increase of the inclination angle will lead to higher radiation intensities when the impact parameter does not exceed $2h_{eff}$. The critical angle for $\ell=10$ is rather small, but it can be scaled by transition to a large OAM. In the case of $\ell=100$, we can expect a strong manifestation of the quadrupole contribution, observing flat behavior of the radiation intensity for the inclination angles from 0 to $0.06 \deg$   

\section{Conclusions} \label{sec:conclusions}
 We have calculated Smith-Purcell radiation generated by a free electron wave packet with an electric quadrupole moment in the range from THz to optical frequencies. 
 Diffraction grating is assumed to be ideally conducting, and arbitrary inclination angles of the wave packet passage are considered. The state of an electron can be described by the Gaussian-Laguerre wave packet with an orbital angular momentum or a Gaussian wave packet with an axial symmetry only. In the former case, the non-paraxial parameter is scaled with the OAM, while in the latter case, it depends on the aspect ratio of the wave packet. 
 In the case of a Gaussian wave packet, our results are exact within the quasi-classical approximation.

As the electron moves near the grating, the spreading of the packet leads to an increase of the quadrupole moment, which also can be seen in the radiation. Although the quadrupole radiation is low as long as the multipole expansion stays valid (in practice, almost always), it leads to an interesting effect: while the radiation intensity from a charge linearly increases with the number of the grating periods, the quadrupole contribution leads to a faster qubic growth, which resembles the coherence effect (superradiance) from a classical many-particle beam. However, in our problem, \textit{ this is a purely quantum effect of the spatial coherence of a vortex packet}. For relativistic particles, spreading can be neglected,  but for non-relativistic and even moderately relativistic electrons (with kinetic energies of ~ 100-300 keV), this effect can lead to both a change in the angular distribution and an increase in the total radiation loss. 

Note that a large OAM leads to a quick spreading and requires short gratings, while small OAM results in a relatively slow spreading and allows using longer gratings. Nevertheless, in the case of inclined passage, we can assume that the wave packets have a large OAM and small Rayleigh lengths.   

We have shown a possibility to track the effects of spatial coherence of the wave packets with intrinsic angular momentum by measuring the intensity of radiation with varying inclination angle (Fig. \ref{fig:inclination-dependence}). The observed dependencies of 
the radiation intensity reveal a qualitative difference between the cases of a charge and a structured wave packet. The radiation of a charge demonstrates exponential decrease with the inclination angle. In contrast, the radiation from a wave packet increases when $\varphi_I\in (0,0.5\varphi_c)$ and the impact parameter $h<2 h_{eff}$. The critical angle is defined by the Rayleigh length, and in the case of a vortex electron, it is proportional to the OAM.   
Our calculations show that experimental observations of the quadrupole contribution to the S-P radiation can be done with a moderate value of OAM, $\ell\sim 100$, or with Gaussian packets with large asymmetry.  The reverse is also true: an inverse problem can be used to determine the wave packet ``shape'' by observing the resulting Smith-Purcell radiation, which provides a non-invasive experimental tool to study, for instance, 
vortex electrons.

\section*{Acknowledgements} \label{sec:acknowledgements}
 We are grateful to A.A. Tishchenko and P.O. Kazinski for fruitful discussions. The study in Sections 2,3 is supported by the Russian Science Foundation (Project No. 17-72-20013). The studies in Section 4 are supported by the Government of the Russian Federation through the ITMO Fellowship and Professorship Program.

\section{Appendix} \label{sec:appendix}
\subsection{Electromagnetic fields of LG wave packet in the rest frame}  
 
Consider a vortex electron described by the LG packet with $n=0$ or an axially symmetric Gaussian wave packet with a non-unity aspect ratio. Its electromagnetic fields are a sum of those of the charge $e$ and of the electric quadrupole moment $Q_{\alpha\beta}$, \eqref{axially-sym-q-moment} (and of the magnetic moment ${\bm \mu}$ in the case of vortex electrons). The fields in cylindrical coordinates in the rest frame were calculated in \cite{karlovets2019dynamical}. In our problem, we prefer to use the Cartesian coordinates: 
\begin{eqnarray}
&& 
E_x=\frac{x}{r^3}\left[1+\frac{1}{4}\left(\frac{3 Q_0}{r^2}\left(1-5\frac{z^2}{r^2}\right)+Q_2\left(\frac{3t^2}{r^2}\left(1-5\frac{z^2}{r^2}\right)+3\frac{z^2}{r^2}-1\right)  \right)\right],
\cr
&& 
E_y=\frac{y}{r^3}\left[1+\frac{1}{4}\left(\frac{3Q_0}{r^2}\left(1-5\frac{z^2}{r^2}\right)+Q_2\left(\frac{3t^2}{r^2}\left(1-5\frac{z^2}{r^2}\right)+3\frac{z^2}{r^2}-1\right)  \right)\right],
\cr
&& 
E_z=\frac{z}{r^3}\left[1+\frac{1}{4}\left(\frac{3Q_0}{r^2}\left(3-5\frac{z^2}{r^2}\right)+Q_2\left(\frac{3t^2}{r^2}\left(3-5\frac{z^2}{r^2}\right)+3\frac{z^2}{r^2}-1\right)  \right)\right],
\cr
&&
H_x = \frac{z}{r^5} \left(3x \frac{l}{2m} - \frac{3}{2} Q_2ty\right),
\cr
&&
H_y = \frac{z}{r^5} \left(3y \frac{l}{2m} - \frac{3}{2} Q_2 tx\right),
\cr
&&
H_z = \frac{l}{2 m}\left(3 \frac{z^2}{r^2} - 1\right)\frac{1}{r^3}
\label{F-cartesian-comp}
\end{eqnarray}
where in the case of LG wave packet, we can just substitute $Q_0=\bar\rho_0^2$,
$Q_2=\frac{l^2 \lambda_C^2}{\bar{\rho}_0^2}$, and in the case of a Gaussian packet, we take $\ell=0$ and 
$$Q_0=(\sigma_\perp^2-\sigma_z^2)\,,$$
$$Q_2=(\sigma_\perp^2-\sigma_z^2)\frac{\lambda_c^2}{4\sigma_\perp^2\sigma_z^2}.$$
We now transform these fields to the laboratory frame, where the particle moves along the $z$ axis with a velocity $\langle u \rangle \equiv \beta$ according to the law
$$
\langle z\rangle = \beta t
$$

Applying the Lorentz transformations, we obtain electric fields in the laboratory frame 
\begin{eqnarray}
&& \displaystyle E^{(\text{lab})}_{x} = \gamma  (E_{x} + \beta H_{y}),\cr
&& \displaystyle E^{(\text{lab})}_{y} = \gamma (E_{y} - \beta H_{x})\,,\ E^{(\text{lab})}_z = E_z
\label{FLabtrans}
\end{eqnarray} 
Simultaneously, we need to transform the coordinates and the time as follows \footnote{Note that Ref. \cite{karlovets2019dynamical} treats the fields at a distant point only, which simplifies the Lorentz transformations of angular variables. Here, we use the general formulas.}:
\begin{align}
 {\bm \rho} = \{x,y\} = \text{inv},\ z\rightarrow  \gamma(z - \beta t)=:R_z,\quad t \rightarrow \gamma(t-\beta z)=:T_z ,\cr
 r^2  \rightarrow  \left(\rho^2 +  \gamma^2(z - \beta t)^2\right)\,,\quad \gamma=(1-\beta^2)^{-1/2}
\label{trans}
\end{align} 
We omit the magnetic fields, because to calculate the surface current in the following, we need only the electric field.

\subsection{Fourier transform of the fields \label{subsec:fourier-integrals}}

First, we calculate the Fourier transform of the electric fields produced by the wave packet 
$$
{\bm E}(q_x, y', z',\omega) = \int\limits dx dt\, {\bm E}({\bm r}',t) e^{i\omega t - i q_x x}
$$
in the coordinate system $(x', y', z')$. Note that $x'=x$. Thus, we obtain the same integrals as in \cite{pupasov2021smith}. 
Consider the general structure of these integrals   
\begin{align}
{\bm E}(q_x, y', z',\omega) =\int\limits dxdt\, {\bm f}(x,y',z')\frac{R_{z'}^{n}}{R'^{\frac{\nu}{2}}} e^{i\omega t - iq_x x}=\gamma^n{\rm e}^{i\omega z'/\beta} f(i\partial_{q_x},y', z')(i\beta\partial_\omega)^{n} {\rm e}^{-i\omega z'/\beta}I_\nu(q_x,y',z',\omega)\,,
\label{fourie-transform-structure}
\end{align}
where $R_{z'}=\gamma(z'-\beta t)$ and ${\bm f}(x,y',z')$ are  polynomial functions depending on $x$, $y'$ variables (the maximum degrees of $x$,$y'$ are equal to 1, and the maximum $z'$ degree is 2 for the quadrupole contribution).
Note that the master integrals 
\begin{eqnarray}\label{def:master-integrals}
I_{2n+1}(q_x,y',z',\omega)=
\int\limits_{-\infty}^{\infty}dt \int\limits_{-\infty}^{\infty}dx \frac{{\rm e}^{i(\omega t-q_x x)}}{\left(x^2+y'^2+\gamma^2(z'-\beta t)^2\right)^{(2n+1)/2}}=\cr
\frac{2\pi}{\gamma\beta}
\frac{i^{n+1}\mu^n h_{n-1}^{(1)}(i\mu|y'|)}{(2n-1)!!|y'|^{n-1}} \exp\left(\frac{iw z'}{\beta}\right)\,,\qquad n=1,2,3\,.
\end{eqnarray}
where $\mu=\sqrt{\frac{\omega^2}{\gamma^2\beta^2}+q_x^2}$, and $h_{n-1}^{(1)}$ are spherical Hankel functions of the first kind  \cite{abramowitz1964handbook}, which contain $y'$ in the denominator.
Nevertheless, the final expressions for the Fourier-transformed fields are polynomials in $y'$, $z'$ and $\text{sgn}(y')$ (up to an exponential factor). It follows from the fact that the only singularity of fields $1/R$ has spherical symmetry.
Such a structure is preserved after the coordinate substitution \eqref{def:coordinate-transformation}. 

In particular, the electric field from the charge 
\begin{align}
&
{\bm E}_e(q_x, y', z',\omega) = \gamma 
\left( i \partial_{q_x},y,\left(z'+i \beta \partial_\omega\right) \right)\,I_3(q_x,y',  z',\omega)\,,\label{Ee-by-master-int}
\end{align}
after the differentiation reads
\begin{align}
&  {\bm E}_e(q_x, y', z',\omega) = \frac{2\pi}{\beta}  \left(-iq_x, \text{sgn}(y') \sqrt{\left(\frac{\omega}{\beta \gamma}\right)^2 + q_x^2}, -i\frac{\omega}{\beta\gamma^2}\right)\,\frac{\exp\left\{i z'\frac{\omega}{\beta} - |y'| \sqrt{\left(\frac{\omega}{\beta \gamma}\right)^2 + q_x^2}\right\}}{\sqrt{\left(\frac{\omega}{\beta \gamma}\right)^2 + q_x^2}}\,.
\label{Eqx}
\end{align}
Calculations of \eqref{fourie-transform-structure} for the quadrupole fields can be done with the aid of computer algebra (see the public repository \cite{pupasov2019git}).  
In the case of the inclined passage, we have to rotate the fields and change the coordinates before the integration along the grating. 
Due to the polynomial dependence on the $y'$, $z'$ variables, 
\begin{align}
 {\bm E}_Q(q_x,y',z',\omega) = 
e^{\left(i z' \left(\frac{\omega\cos(\varphi_I)}{\beta}+i\sin(\varphi_I)\mu\right)\right)}
\left(\tilde {\bm E}_{Q_0}(q_x,y',\omega)+\tilde {\bm E}_{Q_1}(q_x,y',\omega)z'+\tilde  {\bm E}_{Q_2}(q_x,y',\omega)z'^2\right),
\label{E-quadrupole-tildez-polynomial}
\end{align}
Linear coordinate transformation \eqref{def:coordinate-transformation} to $y$ and $z$ variables preserves this polynomial structure. Once again, the quadrupole contribution contains constant, linear and quadratic terms in $z$ variable:
\begin{align}
 {\bm E}_Q(q_x, y, z,\omega) = 
e^{\left(iz \left(\frac{\omega\cos(\varphi_I)}{\beta}+i\sin(\varphi_I)\mu\right)\right)}
\left({\bm E}_{Q_0}(q_x, y,\omega)+{\bm E}_{Q_1}(q_x, y,\omega)z+{\bm E}_{Q_2}(q_x, y,\omega)z^2\right),
\label{E-quadrupole-z-polynomial}
\end{align}
where a $z$-dependent plane wave is multiplied by a second-order polynomial in $z$-variable, where the coefficients are some functions.
The terms linear and quadratic in z contain the quadrupole contribution only and represent the non-paraxial contributions mentioned earlier. 
   
\subsection{Heisenberg equations for the quadrupole moments}  
 Let us derive a general equation that describes the time dependence of 
 the electric quadrupole momentum of a free scalar non-relativistic packet $\psi ({\bm r},t)$.
 Starting with the components of the current $j^{\mu} = \{j^0, {\bm j}\}$, which are
\begin{eqnarray}
& \displaystyle j^0 ({\bm r},t) = |\psi ({\bm r},t)|^2,\ \int d^3 r\, j^0 ({\bm r},t) = 1,\cr 
& \displaystyle {\bm j} ({\bm r},t) = \psi^*({\bm r},t)\frac{-i}{2m}{\bm\nabla}\psi({\bm r},t) + \text{c.c.}
\label{j}\,,
\end{eqnarray}
we define the first three multipole moments corresponding to the initial state $\langle\psi|$, in the Heisenberg picture  
\begin{eqnarray}
& \displaystyle {\bm d}(t) =\langle \psi| {\bm r}(t)|\psi\rangle ,\ \ 
{\bm \mu}(t) = \frac{1}{2m} \langle \psi| {\bm r}(t) \times {\bm p}(t)|\psi\rangle,\cr
& \displaystyle Q_{\alpha\beta}(t) =  \langle \psi|\left (3 r_{\alpha}(t)r_{\beta}(t) - {\bm r}^2(t) \delta_{\alpha\beta}\right)|\psi\rangle,
\label{dmu}
\end{eqnarray}
and the corresponding intrinsic values \cite{karlovets2018relativistic,karlovets2019dynamical}
\begin{eqnarray}
& \displaystyle {\bm d}_{\text{int}} = 0,\ {\bm \mu}_{\text{int}}(t) = {\bm \mu}(t) - \frac{1}{2m}{\bm d}(t) \times \langle \psi| {\bm p}(t)|\psi\rangle ,\cr
& \displaystyle Q_{\alpha\beta,\text{int}}(t) = Q_{\alpha\beta}(t)-3d_{\alpha}(t)d_{\beta}(t) + {\bm d}^2(t) \delta_{\alpha\beta},\cr 
& \displaystyle \alpha,\beta = 1,2,3.
\label{dmuint}
\end{eqnarray}
Only the intrinsic moments are considered throughout the paper, therefore the subscript ``int'' is omitted.
The Hamiltonian of a free particle $H={\bm p}^2/(2m)$ defines the equations of motion $i\partial_t A=[H,A]$
and well-known solutions 
$$
{\bm p}(t)={\bm p}\,\qquad {\bm r}(t)={\bm r}+\frac{{\bm p}}{m}t
$$
Hamiltonian is quadratic in the momentum operator, and the quadrupole moment is quadratic in position operator. From Heisenberg equations, it follows that the third derivative of the quadrupole moment vanishes (it was obtained by direct calculations for the LG packet in \cite{karlovets2019intrinsic}), thus, one can see that
\begin{align}
 Q_{\alpha\beta,\text{int}}(t) =Q_{\alpha\beta,\text{int}}(0)+Q_{\alpha\beta}^{(2)}t^2  
\end{align}

\end{document}